# Placing the Sun in Galactic Chemical Evolution: Mainstream SiC Particles

D. D. Clayton & F. X. Timmes[1]

clayton@gamma.phys.clemson.edu & fxt@burn.uchicago.edu

Department of Physics and Astronomy
Clemson University
Clemson, SC  29634

[1]Present Address:
Astronomy and Astrophysics
UC Santa Cruz
Santa Cruz, CA  95064

ABSTRACT

We examine the consequences and implications of the possibility that the best-fit $m=4/3$ line of the silicon isotopic ratios measured in mainstream SiC grains is identical or parallel to to the mean ISM evolution line of the silicon isotopes. Even though the mean ISM evolution proceeds along a line of unity slope when deviations are expressed in terms of the native representation (the mean ISM), the evolution line can become a slope $4/3$ line in the solar representation, provided that the solar composition is displaced from the mean ISM evolution. During the course of this analysis, we introduce new methods for relating the solar composition to that of the mean ISM at the time of solar birth. These new developments offer a unique view on the meaning of the mainstream SiC particles, and affords a new way of quantitatively answering the question whether the sun has a special composition relative to the mean ISM at solar birth. If the correlation slope of the silicon isotopes in the mean ISM could be decisively established, then its value would quantify the difference between the solar and mean ISM silicon abundances. Our formalism details the transformations between the two representations, and applies not only to $^{29}$Si and $^{30}$Si, but to any two purely secondary isotopes of any element (O, Ne, Mg, and perhaps S). Both the advantages and disadvantages of this technique are critically reviewed.

Subject headings: ISM: Abundances – Nuclear Reactions, Nucleosynthesis, Abundances – Supernovae: General

## 1. INTRODUCTION

A basic and protracted question for astronomy is whether the composition of the sun should be regarded as typical. The degree to which the sun may have an average set of heavy element abundances is evident from consideration of a few examples. Edvardsson et al. (1993) have shown that there is a factor of two scatter in their unbinned [Fe/H] measurements, with slightly smaller amounts of scatter in their unbinned element abundance ratios. They claim that most, if not all, of the scatter is intrinsic to the stars surveyed, and not due to observational uncertainties. Meyer et al. (1994) observed the weak interstellar O I] $\lambda 1356$ absorption feature in the low-density sight lines toward $\iota$ Ori and $\kappa$ Ori. They derived a total oxygen abundance (gas + grains) towards $\iota$ Ori and $\kappa$ Ori that is consistent with stellar and nebular determinations in Orion (Rubin et al. 1991ab; Osterbrock, Tran & Veilleux 1992) at a level that is about 1/2 of the solar oxygen abundance. Snow & Witt (1995) recently analyzed the published data on carbon abundances in field dwarfs, and suggested carbon is more abundant in the sun (350–470 carbon atoms per $10^6$ hydrogen atoms) than in the other stars (225 C atoms per $10^6$ H atoms). These disparate examples give grounds to suspect an atypical abundance of heavy elements in the sun in comparison with other stars that formed $\simeq 4.6$ Gyr ago at a Galactocentric radius of $\simeq 8.5$ kpc. But because it is difficult to know what other set of abundances to take, Galactic chemical evolution models of the mean interstellar medium (henceforth ISM) have generally attempted to reproduce the solar abundances at that time and location.

Hesitation on taking the solar abundances as representative of the mean ISM is perhaps even more acute when isotopic ratios are considered, since less observational information exists for isotopic abundances. Many have questioned whether the solar $^{12}C/^{13}C$ number ratio of 89 (Anders & Grevesse 1989) was typical of its time and place. Averages over the local ISM gas phase abundances currently range between 60-80 (Wilson & Rood 1994; Henkel et al. 1995; Kahane 1995), while cool carbon and J-type stars in the local neighborhood range between 15-60 (Abia & Isern 1996). Many have suggested that either homogeneous chemical evolution has decreased the solar $^{12}C/^{13}C$ number ratio to the local ISM values, or that local inhomogeneities seeded the presolar nebula with an anomalously large $^{12}C/^{13}C$ ratio. A good discussion of both the carbon and oxygen isotopic evolution is given by Prantzos, Aubert & Audouze (1996), in which they repeatedly point to the uncertainty caused by not knowing whether to take solar isotopic abundances as typical.

In this paper we introduce new methods for relating the solar composition to that of the mean ISM at solar birth. The discovery of presolar refractory silicon carbide (henceforth SiC) grains preserved within meteorites has allowed their isotopic compositions to be measured with an unprecedented accuracy (parts per thousand) for astronomy (Virag et al. 1992; Anders & Zinner 1993; Hoppe et al. 1994ab, 1995, 1996). These works have show that the silicon compositions of the grains are not random, and are not scattered over a wide range of values. Instead, the silicon isotopic abundance ratios correlate within a narrow band in a three-isotope plot. The width of the band is $\simeq 20\%$, but the larger grains are generally measured to an unprecedented 0.1% accuracy. This band has been denoted the one in which "mainstream" SiC grains populate, to distinguish them from other SiC grain populations. The mainstream band follows this rule: deviations of the $^{29}$Si/$^{28}$Si ratio from the solar ($^{29}$Si/$^{28}$Si)$_\odot$ ratio are proportional to deviations of the $^{30}$Si/$^{28}$Si ratio from the solar ($^{29}$Si/$^{28}$Si)$_\odot$ ratio. The best-fit slope of the correlation band is 1.35 (Hoppe et al. 1995b), which we hereafter call 4/3.

Clayton (1988) argued that the $^{29}$Si/$^{28}$Si and $^{30}$Si/$^{28}$Si number ratios increase linearly with time in the ISM of the disk, because $^{29}$Si and $^{30}$Si nuclei are both "secondary" nucleosynthesis products, whereas $^{28}$Si is "primary". By "secondary" we mean stellar yields that are proportional to the initial metallicity of the stars that produce them (massive stars, in the case of $^{29}$Si and $^{30}$Si), while by "primary" we mean isotopic yields that are independent of the metallicity. Homogeneous chemical evolution calculations which use the Woosley & Weaver (1995) or the Thielemann, Hashimoto & Nomoto (1996) supernova surveys show that the linear increase of of $^{29}$Si/$^{28}$Si and $^{30}$Si/$^{28}$Si with time is true to a high degree of accuracy (Timmes & Clayton 1996).

In the simplest case in which there is no further isotopic processing by stars, the presolar SiC grains are linearly aligned in a three-isotope plot because the initial compositions of the low-mass stars that made the grains were initially aligned (Clayton 1988). That is, different low-mass stars formed at different times, and thus had different silicon isotopic ratios in their ejecta. If this were the case, the observed SiC grain correlation line would measure the displacement, if any, between solar silicon isotopic ratios and that of the mean ISM. While the observed mainstream SiC grains have a correlation slope of $m=4/3$ and the mean ISM has a m=1 slope [1], several complications render the approach of attributing this difference directly to differences between the presolar nebula and the mean ISM difficult.

The main complication is that SiC grains bear distinct nucleosynthetic signatures of having evolved through the carbon star phase of the asymptotic giant branch (henceforth AGB). Since additional isotopic processing occurs, the simplest one-step model described above is not strictly valid (see Clayton 1988; Clayton, Scowen & Liffman 1989; Alexander 1993). Various approaches have looked at how the AGB evolution may alter the silicon isotopes in such a way that the mean chemical evolution m=1 slope line is systematically steepened to a $m=4/3$ slope line. Timmes & Clayton (1996) attributed the steepening to a systematic effect by which AGB stars with a smaller initial metallicities have a larger fraction of their helium-shell mixed into their envelope during the ejection and condensation process than do AGB stars with a larger initial metallicity. In applying this model, they strongly suggested that mean chemical evolution calculations should be made to pass *exactly* through solar silicon. Tantamount to an adjustment of the supernova yields, they termed this procedure as a "renormalization". Gallino et al. (1994) argued that AGB stars of differing initial metallicity develop different neutron fluences, and thereby steepen the inherited m=1 slope line into a $m=4/3$ slope line. Their mean ISM $^{29}$Si and $^{30}$Si evolution curves are quite different (and probably incorrect) from the evolution curves in the Timmes & Clayton study. The $^{29}$Si/$^{28}$Si and $^{30}$Si/$^{28}$Si ratios evolve down the m=1 evolution line (hence backward in time) in the

---

[1] An italic case $m$ is reserved for slopes measured with respect to solar abundances, while a roman case m is reserved for slopes measured with respect to the mean ISM abundances.

Gallino et al. model, while they evolve up the m=1 line (hence forward in time) in the Timmes & Clayton model.

There is another complication of attributing the difference between the measured SiC grain correlation slope and the mean ISM correlation slope directly to differences between the presolar nebula and the mean ISM. The mainstream SiC grains are richer in the heavier silicon isotopes than the sun, even though the grains must have originated in stars that formed prior to the sun. Homogeneous chemical evolution curves are monotonic and simply cannot accommodate that experimental fact. Timmes & Clayton speculated that the presolar nebula was simply deficient in $^{29}$Si and $^{30}$Si, whereas Gallino et al. removed this causality problem altogether by speculating that the chemical evolution of the mean ISM progressed down the three-isotope line rather than up the line.

The renormalized mean ISM evolution of Timmes & Clayton (1996), and how it progresses up the m=1 slope line with time, is shown in the three-isotope plot of Figure 1. This evolution line has its deviations expressed with respect to the calculated silicon isotopic composition at solar birth (note subscripts on axes labels, and see eq. 2). Silicon isotopic compositions for some of the available Murchison SiC grain data (177 data points; Hoppe et al. 1995b) is also shown in Fig. 1, the full set (584 data points; Hoppe et al. 1995b, 1996) being suppressed simply for clarity. These SiC grains have a best fit slope of 4/3, and have their deviations expressed with respect to the solar isotopic composition (see eq. 2). Deviations expressed with respect to mean ISM and solar abundances are the same, $\delta_\odot = \delta_{\rm ISM}$, whenever renormalization ensures that the mean ISM evolution passes exactly through solar abundances at t=$t_\odot$ (as is the case in Fig. 1). We place this figure here in order to clarify the observational situation, and the alternative solutions that we will later pursue.

Although presolar SiC grains have something important to say about the relative position of sun and the mean ISM, the complications summarized above shroud their message. The physical picture needs further clearing up before the answer can be discerned, and a systematic expansion of the possibilities seems a desirable route. In this paper, we present physical grounds for altered evolution lines based on the finding that the mean ISM chemical evolution line of slope m=1 becomes rotated when deviations are expressed with respect to solar abundances (Timmes & Clayton 1996). The more the mean ISM silicon isotopic ratios differ from the solar silicon isotopic ratios, the greater the degree of rotation. We now analyze the mathematical basis for this interesting situation.

## 2. THE TRANSFORMATION

Silicon isotopic ratios are conventionally expressed in parts per thousand deviation from the solar silicon isotopic ratios:

$$\delta_\odot \left( \frac{^{29}{\rm Si}}{^{28}{\rm Si}} \right) \equiv 1000 \left[ \left( \frac{^{29}{\rm Si}}{^{28}{\rm Si}} \right) \bigg/ \left( \frac{^{29}{\rm Si}}{^{28}{\rm Si}} \right)_\odot - 1 \right]$$
$$\delta_\odot \left( \frac{^{30}{\rm Si}}{^{28}{\rm Si}} \right) \equiv 1000 \left[ \left( \frac{^{30}{\rm Si}}{^{28}{\rm Si}} \right) \bigg/ \left( \frac{^{30}{\rm Si}}{^{28}{\rm Si}} \right)_\odot - 1 \right] \quad . \tag{1}$$

These will be compactly denoted as $\delta_\odot^{29}\text{Si}$ and $\delta_\odot^{30}\text{Si}$, respectively. The subscript serves as an explicit reminder that these deviations are with respect to the composition of solar silicon. The factor of 1000 enters because isotopic anomalies within presolar particles are conventionally expressed in parts per thousand. If the deviations are read as number fractions instead, the factor of 1000 may be omitted. For completeness, the numerical values of the solar silicon isotope ratios are $(^{29}\text{Si}/^{28}\text{Si})_\odot = 0.05253$, and $(^{30}\text{Si}/^{28}\text{Si})_\odot = 0.03599$ (Anders & Grevesse 1989).

Silicon isotopic ratios may also be expressed in parts per thousand deviation from the mean ISM silicon isotopic ratios:

$$\delta_{\text{ISM}}\left(\frac{^{29}\text{Si}}{^{28}\text{Si}}\right) \equiv 1000\left[\left(\frac{^{29}\text{Si}}{^{28}\text{Si}}\right)\bigg/\left(\frac{^{29}\text{Si}}{^{28}\text{Si}}\right)_{\text{ISM}} - 1\right]$$
$$\delta_{\text{ISM}}\left(\frac{^{30}\text{Si}}{^{28}\text{Si}}\right) \equiv 1000\left[\left(\frac{^{30}\text{Si}}{^{28}\text{Si}}\right)\bigg/\left(\frac{^{30}\text{Si}}{^{28}\text{Si}}\right)_{\text{ISM}} - 1\right] \;, \tag{2}$$

and will be compactly denoted as $\delta_{\text{ISM}}^{29}\text{Si}$ and $\delta_{\text{ISM}}^{30}\text{Si}$, respectively. This subscript functions as a reminder that the deviations are with respect to the mean ISM in the solar neighborhood at the time $t_\odot$ when the sun formed. Precisely what the value of $t_\odot$ is relative to the age of the Galaxy is not well determined; but this is not a serious limitation. We simply focus on the mean ISM at that time, which was $4.6 \pm 0.1$ Gyr ago according to various nuclear chronometers (Wasserburg et al. 1977).

A linear coordinate transformation connects the two reference bases:

$$\begin{aligned}\delta_\odot^{29} &= 1000\left[\frac{^{29}\text{Si}/^{28}\text{Si}}{(^{29}\text{Si}/^{28}\text{Si})_\odot} - 1\right] \\ &= 1000\left[\frac{^{29}\text{Si}/^{28}\text{Si}}{(^{29}\text{Si}/^{28}\text{Si})_{\text{ISM}}} \cdot \frac{(^{29}\text{Si}/^{28}\text{Si})_{\text{ISM}}}{(^{29}\text{Si}/^{28}\text{Si})_\odot} - 1\right] \\ &= 1000\left[\frac{(^{29}\text{Si}/^{28}\text{Si})_{\text{ISM}}}{(^{29}\text{Si}/^{28}\text{Si})_\odot} \cdot \left(\frac{\delta_{\text{ISM}}^{29}}{1000} + 1\right) - 1\right] \;.\end{aligned} \tag{3}$$

It is useful from a physical viewpoint, as well being a notation compactifier, to define a quantity for the silicon isotope ratio in the mean ISM at $t_\odot$ to the solar silicon isotope ratio. By letting

$$\theta^{29} = \frac{(^{29}\text{Si}/^{28}\text{Si})_{\text{ISM}}}{(^{29}\text{Si}/^{28}\text{Si})_\odot} \;, \tag{4}$$

eq. (3) simplifies to the equation of a straight line with slope $\theta^{29}$ and intercept $1000\,(\theta^{29} - 1)$:

$$\delta_\odot^{29} = 1000\left[\theta^{29}\left(\frac{\delta_{\text{ISM}}^{29}}{1000} + 1\right) - 1\right] = \theta^{29}\delta_{\text{ISM}}^{29} + 1000\,(\theta^{29} - 1) \;. \tag{5}$$

Similar manipulation for $\delta_\odot^{30}$ produces

$$\delta_\odot^{30} = 1000\left[\theta^{30}\left(\frac{\delta_{\text{ISM}}^{30}}{1000} + 1\right) - 1\right] = \theta^{30}\delta_{\text{ISM}}^{30} + 1000\,(\theta^{30} - 1) \;, \tag{6}$$

where, in analogy with eq. (4),

$$\theta^{30} = \frac{(^{30}\text{Si}/^{28}\text{Si})_{\text{ISM}}}{(^{30}\text{Si}/^{28}\text{Si})_\odot} \quad . \tag{7}$$

Note eqs. (5) and (6) express a point-to-point transformation between solar and mean ISM representations. When these transformations act on a set of linearly correlated points (a line), the transformation represents a translation (the intercept) and a rotation (the slope) in a three-isotope plot. A three-isotope plot refers to coordinate axes involving isotope ratios of the same element with the same isotope appearing in the denominator ($^{28}$Si in the case of Fig. 1). It has proven especially useful owing to mathematical properties for mixtures of compositions. A mixture between any two points on any three-isotope plot must produce a composition lying on the line connecting the two points, with the mix point composition located at the center of mass.

## 3. ROTATED EVOLUTION LINE FOR SECONDARY ISOTOPES

Consider the case when $\delta^{29}_{\text{ISM}}$ and $\delta^{30}_{\text{ISM}}$ correlate with time along an evolution line of slope m=1, as they do for secondary isotopes in homogeneous chemical evolution models (see Fig. 1). By definition this evolution line passes exactly through the origin ($\delta^{29}_{\text{ISM}} = 0$, $\delta^{30}_{\text{ISM}} = 0$) at the time of solar birth t=$t_\odot$. That is, the evolution line for two strictly secondary isotopes in the mean ISM representation is

$$\delta^{29}_{\text{ISM}} = \delta^{30}_{\text{ISM}} \quad , \tag{8}$$

where formally the $\delta$'s are now functions of time. We want to examine how this unity slope evolution line appears when viewed from a solar representation, and proceed as follows. Substituting the specific evolution of eq. (8) into the general transformation between solar a mean ISM representations given by eq. (5), yields

$$\delta^{29}_\odot = \theta^{29}\delta^{30}_{\text{ISM}} + 1000 \ (\theta^{29} - 1) \quad . \tag{9}$$

Substituting for $\delta^{30}_{\text{ISM}}$ from eq. (6) gives

$$\begin{aligned}
\delta^{29}_\odot &= \frac{\theta^{29}}{\theta^{30}} \left[\delta^{30}_\odot - 1000 \ (\theta^{30} + 1)\right] + 1000 \ (\theta^{29} - 1) \\
&= \frac{\theta^{29}}{\theta^{30}} \delta^{30}_\odot + 1000 \left[\frac{\theta^{29}}{\theta^{30}} - 1\right] \quad .
\end{aligned} \tag{10}$$

Finally, by letting

$$m = \frac{\theta^{29}}{\theta^{30}} \quad , \tag{11}$$

eq. (10) becomes the equation of a straight line having slope $m$ and intercept $1000(m-1)$:

$$\delta^{29}_\odot = m \ \delta^{30}_\odot + 1000 \ (m - 1) \quad . \tag{12}$$

Thus, a unity slope evolution in the mean ISM representation (eq. 8) appears in the solar representation as a line that is translated along the $\delta^{29}_\odot$ axis by an amount equal to the intercept.

This transformed line passes through the ($\delta^{29}_\odot$=0,$\delta^{29}_\odot$=0) origin only for the special case $m = 1$. Referenced to the solar basis, the mean evolution line does not have a slope of unity, but a slope given by eq. (11).

Figure 2 shows eq. (12) for several choices of $m$. The shaded band is the region occupied by the mainstream SiC grains (see Fig. 1), which have a best-fit slope of 4/3. Notice that the evolution line for $m=4/3$ lies on the left side of this three-isotope plot whereas the evolution line with $m=2/3$ lies at the right side of the plot.

The mean ISM chemical evolution line of unity became a correlation line of $m \simeq 2/3$ in solar representation when the Woosley & Weaver (1995) supernovae yields were employed (Timmes & Clayton 1996). That is, those supernovae yields evolve to a $^{29}$Si/$^{30}$Si ratio at $t_\odot$ that is only 2/3 of the corresponding solar ratio. Under these conditions eq. (12) becomes

$$\delta^{29}_\odot = \frac{2}{3} \delta^{30}_\odot - 333 \tag{13}$$

which is nearly identical to eq. (13) of Timmes & Clayton. This evolution is shown in Fig. 2. If the supernova yields used are correct, the evolution of the mean ISM must pass to the right and below the solar position, as Timmes & Clayton showed.

The best-fit slope of the mainstream SiC particles is $m=4/3$. One may speculate that the unity slope in the mean ISM representation is transformed into a 4/3 slope in the solar representation. This requires supernovae yields which evolve to a $^{29}$Si/$^{30}$Si ratio at $t_\odot$ that is 4/3 of the corresponding solar ratio. Since the yields of the silicon isotopes from supernovae are uncertain by at least factors of two due, for example, to errors in the nuclear rates, supernovae may well (or may not) drive a mean ISM evolution towards $m=4/3$. Plugging this slope into eq. (12) gives

$$\delta^{29}_\odot = \frac{4}{3} \delta^{30}_\odot + 333 \ . \tag{14}$$

This evolution also shown in Fig. 2. It shows that for this explanation to be relevant, the evolution of the mean ISM must pass to the left and above the solar position.

These examples of eq. (12) illustrate the primary finding of this work: if the correlation slope of the silicon isotopes in the mean ISM could be established, its value would quantify the difference between the solar and mean ISM silicon compositions. It could measure how anomalous solar silicon is. The same procedures and conclusion apply to oxygen, because $^{17}$O and $^{18}$O are also secondary nucleosynthesis products, and probably to Ne, Mg and S as well.

## 4. THE AGE PARAMETER

All silicon isotopic ratio evolutions having a slope $m=4/3$ in the solar representation are represented by a unique line in Fig. 2. This requires, per eq. (11), that the nuclear details of massive star nucleosynthesis produce a $^{29}$Si/$^{30}$Si ratio in the mean ISM at $t_\odot$ that is 4/3 of the solar value. This demands the solar silicon abundances to be anomalous. While the overproduction of $^{29}$Si ($\theta^{29}$ in eq. 4), and $^{30}$Si ($\theta^{30}$ in eq. 7), are not constrained individually, their ratio is constrained to have the ratio 4/3.

The value of $\theta^{29}$ does, however, determine how far along the $m=4/3$ line that the evolution has proceeded when a time of $t_\odot$ has been reached. In this regard, $\theta^{29}$ acts as an age parameter. Each choice of $\theta^{29}$ (filled circles) in Figure 3 reaches a different point on the line at $t_\odot$. The deviations at $t_\odot$, expressed in the solar representation, are given by by setting the ISM deviations to zero in eqs. (5) and (6):

$$\delta_\odot^{29} = 1000\ (\theta^{29} - 1) \qquad \delta_\odot^{30} = 1000\ (\theta^{30} - 1)\ . \qquad (15)$$

For example, when $\theta^{29}=1$ the solar ratio for $^{29}$Si/$^{28}$Si is reached at $t_\odot$ ($\delta_\odot^{29}=0$). In this case $^{30}$Si in the mean ISM must be underproduced by a factor of 3/4 relative to solar. A value of $\theta^{30}=3/4$ is necessary in order that the ratio, given by eq. (11), yields $m=4/3$.

If the sun does not lie on the mean evolution line, then it is presently difficult to say what the precise evolution of nature's mean ISM evolution is. Timmes & Clayton (1996) assumed that the sun does lie on the mean evolution line, and they renormalized the evolution line accordingly. They suggested, in effect, a small systematic error in the supernova yields employed.

Fig. 3 illustrates the problem faced if the $m=4/3$ line for the mainstream SiC grains is to also represent the Galactic evolution of the silicon isotopes. Each star contributing STARDUST to the solar nebula must have been born, evolved, and condensed the SiC grains prior to $t_\odot$ (see discussion in §1). We use the term STARDUST as an acronym for dust that condenses within stellar material being ejected by a star. The initial silicon abundances of these stars are given by the portion of the $m=4/3$ line that lies below the filled circle corresponding to the true value of $\theta^{29}$. Those stars which contributed grains at $t_\odot$ (located at the appropriate filled circle) should, in the simplest case, provide the grains at the tip of the mainstream SiC band (largest $\delta_\odot^{29}$ and $\delta_\odot^{30}$). This is illustrated by the dashed line connecting the $\theta^{29}=1$ evolution point to the tip of band that covers the mainstream SiC grains.

We emphasize this point and dashed line for two reasons: (1) a value of $\theta^{29}=1$ is near the value calculated by Timmes & Clayton (1996) using current supernova yields; (2) this point and the tip of the mainstream SiC band is connected by a slope 1/2 line, which is near the slope expected for s-process modifications of silicon by AGB stars.

## 5. AGB SHIFTS ALONG m=1/2

Under the scenario under consideration, Fig. 3 shows that individual AGB stars must make a large change in their silicon isotopic composition in order to reach the mainstream band. Representative values are easily derived. The grains at the tip of the mainstream band have $\delta_\odot^{29} \simeq 200$ and $\delta_\odot^{30} \simeq 150$. To have all AGB stars which contribute to the mainstream grains born before the sun, one wants $m=4/3$ and (by construction) $\delta_{\rm ISM}^{29} = \delta_\odot^{29} = 0$ at $t=t_\odot$. Inspection of eqs. (4) through (7) shows that all these conditions can be satisfied when

$$\theta^{29} = 1 \qquad \theta^{30} = 3/4$$
$$\delta_\odot^{29} = \delta_{\rm ISM}^{29} \qquad \delta_\odot^{30} = 3/4\ \delta_{\rm ISM}^{30} - 250\ . \qquad (16)$$

That is, AGB stars must move their ISM inherited composition of ($\delta_\odot^{29} = 0$, $\delta_\odot^{30} = -250$) to the top the mainstream at ($\delta_\odot^{29} = 200$, $\delta_\odot^{30} = 150$). These are total changes of ($\delta_\odot^{29}=200$, $\delta_\odot^{29}=400$), which must be provided by modification of the AGB star's initial silicon composition.

Is this large of a modification reasonable? The answer hinges upon when SiC grains form in the evolution of intermediate-mass stars. Dredged-up helium-shell material, which has experienced significant s-processing, has silicon isotopic ratios near ($\delta_\odot^{29}=400$, $\delta_\odot^{30}=900$). This material exceeds the required modification by a factor of two, so the modification is at least feasible. In the early winds of the carbon phase, however, the dredged-up material is severely diluted by the envelope material. Grains which may condense from this diluted material will fail to give a large enough modification to the silicon isotope ratios. But in the final stages of AGB evolution, just prior to reaching the beginning of the white dwarf cooling sequence, there is very little envelope material remaining with which to dilute the dredged-up helium-shell matter. Grains which may condense from this late evolution stage material exceed the required modification by an ample margin.

So, the conjecture given above for the $m=4/3$ slope may succeed if good physical reasons can be established for the mainstream SiC grains being very rich in dredged-up helium-shell material. Pursuing this line of investigation is not the main concern of this paper, but it is quite interesting to note the following. The measured SiC grains are very large for ISM grains, with most of the SiC grains having a linear dimension between 1–5 $\mu$m. Such large grains are not easily grown in AGB winds. For a homologous expansion, the number density of particles in a typical AGB wind decreases as t$^{-2}$. The integral of N(C) and N(Si) over time thus converges, but to values that imply grain sizes $\simeq$ 0.1 $\mu$, even if the the coagulation process operates at 100% efficiency (i.e., only perfect inelastic collisions). Thus, growing large SiC grains requires a silicon-rich and carbon-rich environment that is at a large enough density for a sufficiently long time. Now, there are a few examples of AGB stars thought to be undergoing their final helium flash: Abell 30, Abell 78, Abell 58/V605 Aql (also known as Nova Aquilae 1918), FG Sagittae, R Coronae Borealis, and most recently (and spectacularly) Sakurai's object (Searle 1961; Herbig & Boyerchuck 1969; Langer, Kraft & Anderson 1974; Hazard et al. 1980; Ford & Jacoby 1983; Iben et al. 1983; Duerbeck 1989; Kaler 1989; Duerbeck & Benetti 1996). We will concentrate on a few properties of Sakurai's object since it may be the best documented case, but most of our comments hold for the other planetary nebulae which have a final he-flash core. First, the photospheric radial speed appears identical to the radial speed of the surrounding nebula, indicating that the two are associated, and that there is currently no mass loss by a think wind (Duerbeck & Benetti 1996). The final dredged-up material and the nebula appear to be slowly lifting off in unison at $\simeq$ 25 km s$^{-1}$, based on the nebula's [O III] emission lines. Second, the C I, C II, and Si II lines are extraordinarily strong, indicating the material being ejected by the final helium-flash is quite rich in carbon and silicon (Duerbeck & Benetti 1996). Third, theoretical models assign a mass density to the final ejecta that is at least an order of magnitude greater than the density in the early carbon-star ejecta (Iben et al. 1983; Iben 1984). In this type of carbon and silicon rich environment and at these small lift off speeds, the time integrated number density may be large enough to grow SiC particles in the 1–5 $\mu$m range. To be blunt, the material lifted off by the final helium-flashes may be the ideal place for nature to grow the mainstream SiC grains. If observational evidence is found to favor SiC grain formation from very helium-shell rich AGB material, then the $m=4/3$ slope may have been inherited from the initial AGB star composition, without the need for a skewing of the slope by nuclear effects such as those postulated by Timmes & Clayton (1996) or Gallino et al. (1994).

## 6. CONSTANT STELLAR SHIFT IN STARDUST

Suppose that each star formed on the ISM evolution line ejects STARDUST that has altered the initial isotopic ratios by constant amounts:

$$\delta^{29}_{\text{ISM,dust}} = \delta^{29}_{\text{ISM}} + b^{29} \qquad \delta^{30}_{\text{ISM,dust}} = \delta^{30}_{\text{ISM}} + b^{30} \quad . \tag{16}$$

Because of the unit slope for the mean ISM, giving the condition $\delta^{29}_{\text{ISM}} = \delta^{30}_{\text{ISM}}$ at all times, it follows that

$$\delta^{29}_{\text{ISM,dust}} = \delta^{30}_{\text{ISM,dust}} + b \quad . \tag{17}$$

where $b = b^{29} - b^{30}$. These could be, for example, WR stars or AGB stars changing the initial silicon by constant amounts. One wishes to see how dust bearing silicon that has been altered by stellar evolution appears in the special solar representation. Redoing the preceding algebra shows the dust line, in the solar representation, has a slope $m$ and an intercept $1000\,(m-1) + b\,\theta^{29}$:

$$\delta^{29}_{\odot} = m\,\delta^{30}_{\odot} + 1000\,(m-1) + b\,\theta^{29} \quad . \tag{18}$$

This equation reduces, as it should, to eq. (12) when $b = 0$.

Figure 4 shows eq. (18) for the case $b = -200, \theta^{29} = 0.937$ and several choices of the slope $m$. The choice of $b$ is arbitrary, while the choice of $\theta^{29}$ is the value found by Timmes & Clayton (1996). Note that the values of $b$ in are expressed in their ISM representation; one must not carelessly insert values expressed with respect to solar abundances. For physically derived shifts, which may be more variable than the simple constant case considered above, one calls upon the basic transformation pair (eqs. 1 and 2) and performs the transform numerically.

## 7. SUMMARY

We have presented and analyzed the general relationship between three-isotope evolution lines in the mean ISM representation and the solar representation (eqs. 4 and 6, respectively). This general transformation applies to any pair of strictly secondary isotopes (e.g. oxygen as well as silicon), and in this sense seems to merit open consideration.

We have examined the possibility that the best-fit $m=4/3$ line of mainstream SiC grains can be the same as mean ISM chemical evolution line. Even though the mean ISM evolution proceeds along a line of unity slope, when deviations are expressed in terms of the mean ISM basis, it can become a slope $4/3$ line in the solar basis provided that the solar composition is displaced from the mean ISM evolution (eqs. 12; Figs. 2 and 3). We find this development exciting because it not only offers a possibly new glimpse on the meaning of the mainstream SiC particles, but it also affords a new way of fixing the sun's composition relative to the norm at the time of solar birth. It possibly offers a quantitative answer to the question whether the sun has a special composition.

Unfortunately, we cannot press this method to a decisive answer at this time because the astrophysical requirements are somewhat severe and may not hold up to further investigation. For the method to be correct, the stars creating the STARDUST must move the silicon isotopic composition rather far from their initial values. Namely, the stars must increase their $^{29}$Si/$^{28}$Si and $^{30}$Si/$^{28}$Si ratios by about 20% and 40%, respectively (if the value of $\theta^{29}$ is near unity). Neutron irradiation does indeed produce such an effect. Model-dependent calculations of AGB star evolution suggest that in the helium-shell material dredged up to the surface during the AGB lifetime, the changes due to s-processing are near 40% and 90%, respectively (Brown & Clayton 1992; Gallino et al. 1994; also see Boothroyd, Sackmann & Wasserburg 1995). This explanation requires that the neutron-irradiated silicon isotopes constitute about 1/2 of the silicon isotopes in the SiC grains. To be plausible there must exist an independent reason to expect such a thing. We have tentatively attributed an independent reason to the large size of the SiC grains, by suggesting that during most of the carbon star phase smaller grains are condensed in their winds, and that only during the last pre-planetary-nebula lift-off can grains grow larger than 1 $\mu$m. If this suggestion is correct, we anticipate a band of small SiC grains extending down and to the left of the mainstream SiC grains.

It behooves us to point out further implausibilities. For this scenario to be true, the evolution of the mean ISM must lead to a $^{29}$Si/$^{30}$Si ratio at $t_\odot$ that is larger than solar, despite the fact that current calculations produce values that are smaller than solar. This is hardly what anyone would have guessed, but is by no means impossible, since yields from supernovae are accurate to no better than a factor of two. As Fig. 4 shows, the true ISM at $t_\odot$ would have to be be 25% deficient in $^{30}$Si with respect to solar (if $\theta^{29}$=1). For any values of $\theta^{29}$, it is difficult to understand how the sun can differ from the mean ISM at $t_\odot$ by 25% in its silicon isotopic ratios.

The location of the solar silicon abundances in the three-isotope plot of Figs. 1–4 is uncomfortable, lying near the bottom of the band populated by the mainstream SiC grains. This implies the solar Si isotopes are "less evolved" than the stars whose SiC grains populate the mainstream band. The sun, however, is metal-rich with respect to other 4.5 Gyr old stars in the solar neighborhood (Edvardsson et al. 1993). In canonical stellar-chemical evolution models the secondary isotopes ($^{29}$Si and $^{30}$Si) increase monotonically with time, and hence with metallicity. Evidently, at least one assumption or observation is incorrect.

Wielen et al. (1997) have recently advocated that the sun formed at 6.6 kpc and has diffused outwards to the present 8.5 kpc. The simple version of their argument is that the sun is known to be $\simeq$ 0.2 dex richer in [Fe/H] than nearby stars, and the Milky Way's disk is known to support a radial gradient of [Fe/H] $\simeq -0.1$ dex/kpc. If the gradient has remained in place for the last 4.5 Gyr, then the sun was born $\simeq$ 2 kpc closer to the Galactic center than its present location. This argument clearly requires more detailed justification and Wielen et al. have begun that process. Diffusion, which is to say the scattering of stellar orbits by the grainy and time dependent gravitational potential well of the Milky Way's disk, is no larger for the sun than for any significant fraction of other stars at given radius. Wielen et al. show diffusion can successfully account for the increase in velocity dispersion with stellar age, and the larger dispersion in metallicity of nearby stars with increasing age. Working backwards in time, Wielen et al. show the initial metallicity dispersion at any radius is rather small; that is, homogeneous evolution of the gas appears to be a good approximation. The notion of a movable sun may be an attractive idea for interpretations of mainstream SiC grains, and other STARDUST particles, whenever Galactic evolution effects play a role. It merits study in this regard, but we will not do so here.

Finally, one must notice the surprising coincidence (in this picture) of having a highly anomalous sun lying very close to highly anomalous mainstream SiC particles. Because the sun and the SiC particles differ from true ISM for totally different and presumably independent reasons, it is surprising that they are nearly equal. The sun (in this picture) lies only about 1-2% away from the lower end of the mainstream SiC band.

This work has been supported at Clemson by the W. M. Keck Foundation, by a NASA Planetary Materials and Geochemistry grant (DDC), and by a Compton Gamma Ray Observatory Postdoctoral Fellowship (FXT). We thank Arnold Boothroyd for a pertinent referee report.

# FIGURES AND CAPTIONS

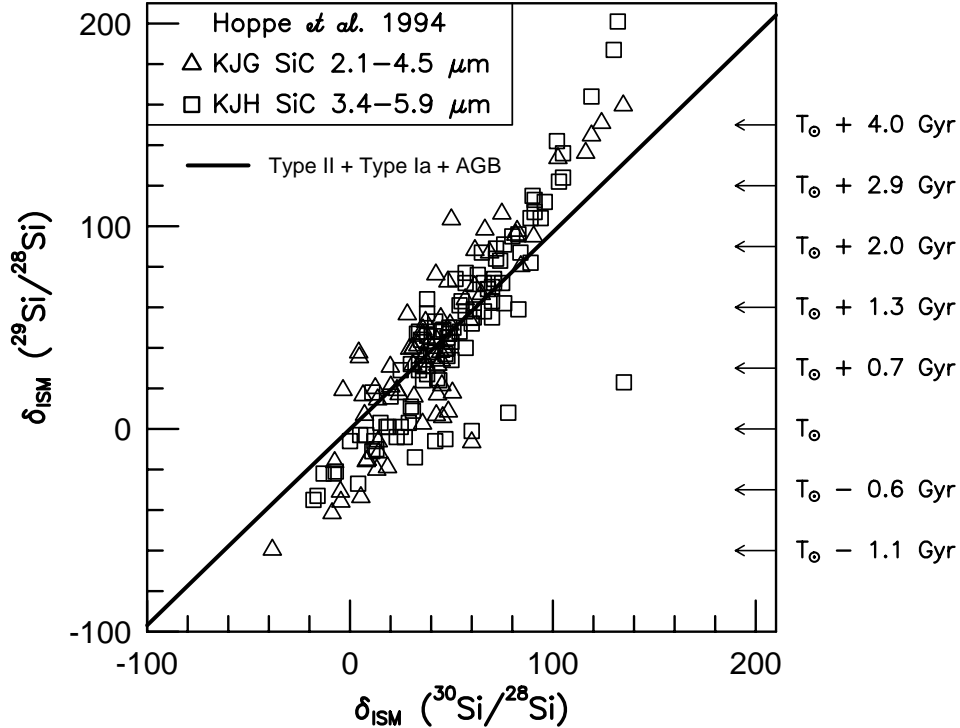

Fig. 1.— Silicon isotope deviations in a three-isotope plot. The renormalized mean ISM evolution of Timmes & Clayton (1996), and how it progresses up the m=1 slope line with time, is shown as the solid line. This evolution line is expressed with deviations expressed with respect to the mean ISM at solar birth (note note subscripts on axes labels and see eq. 2). Chemical evolution models of the mean ISM cannot produce correlation slopes much different than unity, when deviations are expressed in their native ISM representation. Inclusion or exclusion of the Type II supernovae, Type Ia supernovae, and AGB stars are analyzed in detail in Timmes & Clayton. Deviations of the silicon isotopic compositions from the solar composition for Murchison grain data series KJG and KJH (177 data points; Hoppe et al. 1994ab) are shown, the full data set (584 data points; Hoppe et al. 1994ab, 1995, 1996) containing the smaller grain series KJE being omitted for clarity. These SiC grain measurements have a best-fit slope of 4/3 when their deviations are expressed with respect to the solar isotopic composition (see eq. 1). These two representations are the same, $\delta_\odot = \delta_{\rm ISM}$, when silicon isotope evolutions are renormalized to ensure that the mean ISM evolution passes exactly through solar abundances at $t=t_\odot$, as in this figure.

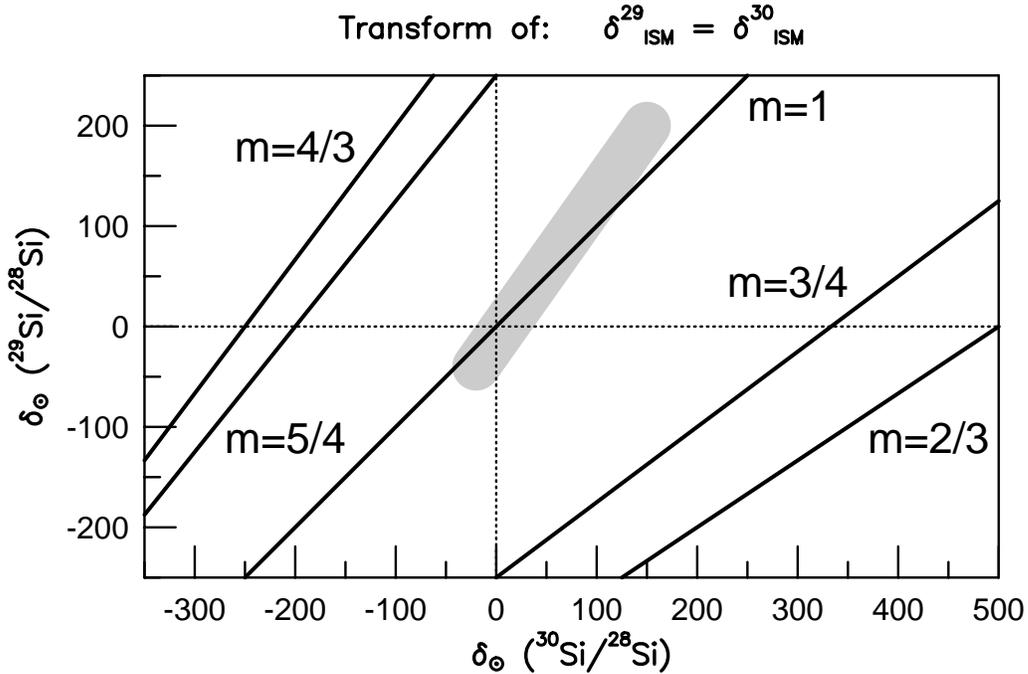

Fig. 2.— Evolution lines of the mean ISM when deviations are expressed relative to solar abundances. Secondary isotopes in mean chemical evolution models correlate along a unit slope evolution line ($\delta^{29}_{\rm ISM} = \delta^{29}_{\rm ISM}$), when deviations are expressed relative to their native mean ISM abundances. This unity slope mean ISM evolution becomes rotated and translated when deviations are expressed with respect to solar abundances. The more the mean ISM silicon isotopic ratios differ from the solar silicon isotopic ratios, the greater the degree of rotation and translation (see eq. 12). Several choices for the difference between the mean ISM and solar silicon abundances, denoted by $m$ in eq. (12), are shown in the figure. The shaded band is the locus of the mainstream SiC grains (see Fig. 1), which have a best-fit slope of 4/3 in the solar representation. The mean ISM line becomes a correlation line of $m=2/3$ when the Woosley & Weaver (1995) supernovae yields are used (Timmes & Clayton 1996). This makes the mean ISM evolution pass to the right and below the solar position. If the mean ISM evolution is transformed into a $m=4/3$ to seek an explanation for the mainstream SiC grains, the evolution of the mean ISM must pass to the left and above the solar position.

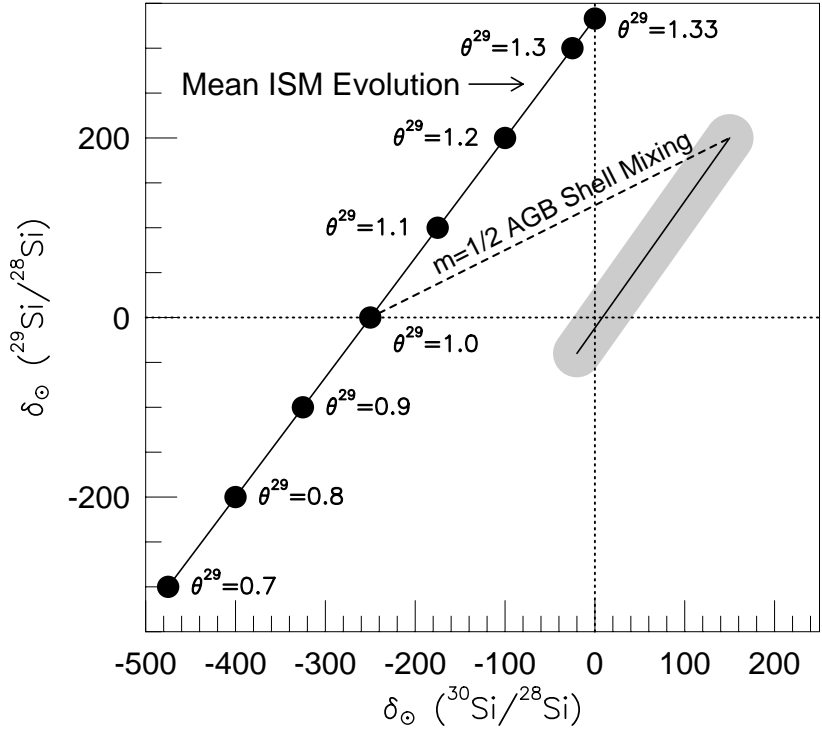

Fig. 3.— Further aspects of the mean ISM evolution line being transformed into a $m=4/3$ slope line in the solar representation. The mainstream SiC grains (shaded band) also have best-fit slope 4/3, but displaced to the right of the transformed mean ISM evolution line. There are no free parameters in the position of the 4/3 line for strictly secondary $^{29}$Si and $^{30}$Si, but the location on the line at the time of solar birth is determined by the value of $\theta^{29}$ (see eq. 15). In this regard, $\theta^{29}$ acts as an age parameter. Each value of $\theta^{29}$ (filled circles) attain a different position along the $m=4/3$ line at the time of solar birth. All AGB stars that contribute to the measured SiC grain population must have been born, evolved, and condensed the grains prior to $t_\odot$. The initial silicon abundances of these AGB stars are on the portion of the $m=4/3$ line that lies below the filled circle corresponding to the true value of $\theta^{29}$. Those AGB stars which contributed grains formed at $t_\odot$ (located at the appropriate filled circle) should, in the simplest case, provide the grains at the tip of the mainstream SiC band (largest $\delta_\odot^{29}$ and $\delta_\odot^{30}$). Now, a $\theta^{29} = 1$ is very near the value (0.937) calculated by Timmes & Clayton (1996) using current supernova yields. In addition, the $\theta^{29} = 1$ point and the tip of the mainstream SiC band are connected by a slope 1/2 (dashed) line, which is near the expected s-process modifications of silicon by AGB stars.

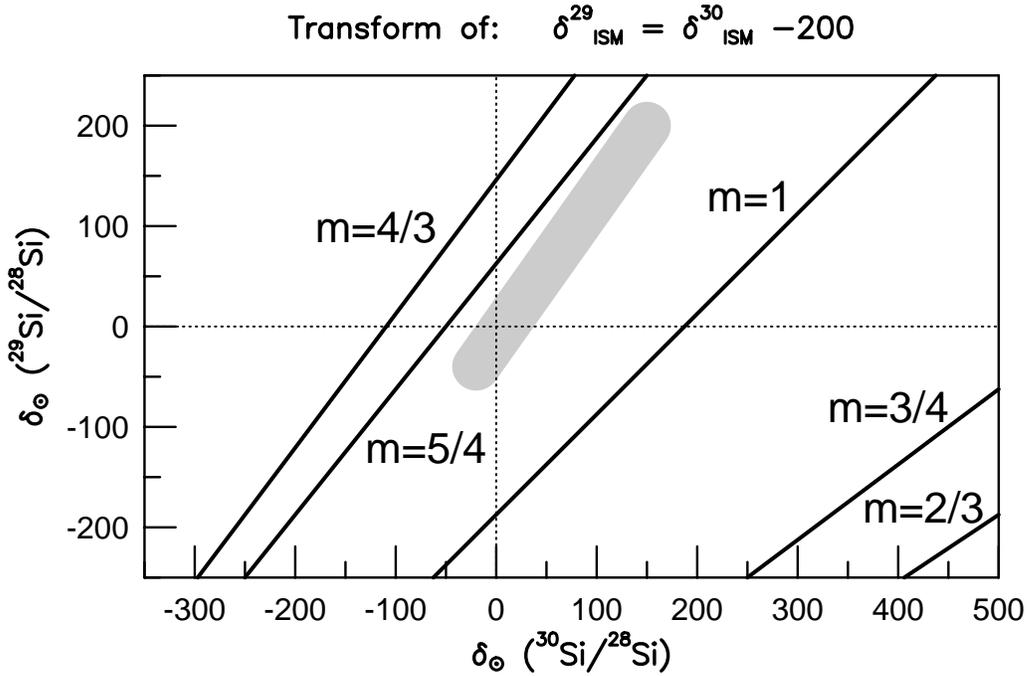

Fig. 4.— Evolution lines of the mean ISM when deviations are expressed relative to solar abundances, and when stars eject grains that have been altered from the initial isotopic ratios by a constant. The figure, similar to Fig. (2), shows eq. (18) for the $b = -200, \theta^{29} = 0.937$ and several choices of the slope $m$. The choice of $b$ in this graph is arbitrary, but is ultimately determined in nature by the displacement appropriate to the stars that create the grains. The choice of $\theta^{29}$ is the value found by Timmes & Clayton (1996), with its true value depending upon the true isotopic yields from supernovae.